\documentclass[aps,prx,twocolumn,groupedaddress,floats,showpacs,final]{revtex4-1}

\usepackage{graphicx}
\usepackage[colorlinks]{hyperref}
\usepackage{makecell}
\usepackage{mathrsfs}
\usepackage{dcolumn}
\usepackage{amsmath}

\begin{document}


\title{Space- and time-crystallization effects in multicomponent superfluids}


\author{Nikolay Prokof'ev}
\affiliation{Department of Physics, University of Massachusetts, Amherst, Massachusetts 01003, USA}
\affiliation{National Research Center ``Kurchatov Institute," 123182 Moscow, Russia}

\author {Boris Svistunov}
\affiliation{Department of Physics, University of Massachusetts, Amherst, Massachusetts 01003, USA}
\affiliation{National Research Center ``Kurchatov Institute," 123182 Moscow, Russia}
\affiliation{Wilczek Quantum Center, School of Physics and Astronomy and T. D. Lee Institute, Shanghai Jiao Tong University, Shanghai 200240, China}
\date{\today}

\date{\today}

\begin{abstract}
We observe that space- and time-crystallization effects in multicomponent superfluids---while having
the same physical origin and mathematical description as in the single-component case---are conceptually
much more straightforward. Specifically, the values of the temporal and spatial periods are absolute rather than relative,
and the broken translation symmetry in space and/or time can be revealed with experiments involving
only one equilibrium sample. We discuss two realistic setups---one with cold atoms and another one with
bilayer superconductors---for observation of space and time crystallization in two-component counterflow
superfluids.

\end{abstract}

\maketitle

The  superfluid long-range order---either genuine or, in lower dimensions, topological/algebraic---is
associated with the emergence of a well-defined (modulo $2\pi$), defect-free field of a coarse-grained phase,
$\Phi({\bf r},t)$; see, e.g., Ref.~\cite{SBP}. In what follows, we discuss the genuine long-range order only.
The generalization to the case of algebraic order is readily achieved along the lines described in
Ref.~\cite{PS_2018}. Also, we use the classical-field (matter-wave) language, which, on one hand,
captures the essence of superfluid phenomena and, on the other hand, is straightforwardly
generalized to the case of quantum bosonic fields \cite{SBP}.

Long-range order in the coarse-grained matter field $\psi = \exp [i \Phi({\bf r},t)]$ means that we are
dealing with the broken global U(1) symmetry state.
The very nature of this state implies the existence of a {\it space crystal} when the phase is a
linear function of distance (and $\psi$ is periodic in space with the period $2\pi/k$):   \\
\begin{equation}
\psi({\bf r},t)\, =\, \psi({\bf 0},t)\, e^{i {\bf k}\cdot {\bf r}} .
\label{plane_wave}
\end{equation}
This is a state with finite superflow velocity proportional to the wavevector ${\bf k}$.
In the literature on superfluidity, the term ``space crystal" is almost never (if at all) used in the context
of the state Eq.~(\ref{plane_wave}), because the matter density $n({\bf r},t)=|\psi({\bf r},t)|^2$ remains
homogeneous in space. [It should not be confused with a supersolid---the superfluid state with
spontaneously broken translation symmetry in the particle density.]
However, it is now conventional to call various states of mater  ``solids" and/or  ``crystals" if there is some
observable revealing broken translation invariance, and this observable need not be the particle density.
One familiar example is the valance-bond crystal state of lattice bosons at half-integer filling factor.
In superfluids, the phase field plays the role of such an observable.
The interference fringes produced by superimposing two matter waves with opposite wavevectors
\cite{Ketterle1995} nicely visualize the fact that superflows break translation symmetry and thus
qualify to be called space crystals. Supercurrent states in three-dimensional superfluids do not form naturally by cooling the system
across the transition temperature \cite{PS_2000}. Nevertheless, supercurrent states can be prepared
by cooling the system in a rotating vessel which is stopped once the system is in the superfluid phase.
In this sense, the period of the space crystal in the phase field depends on the experimental conditions
used to prepare the sample, but otherwise we are dealing with a stable thermodynamic equilibrium 
described by the Gibbs distribution with an emergent quantized topological constant of motion (phase winding number) \cite{SBP}. 
Academically speaking, such a persistent current state is metastable. However, its relaxation time due to 
rare quantum-tunneling or thermal-activation events is exponentially large in the inverse $k$, and
easily exceeds the time of the Universe unless
the period of the space crystal is microscopically small.

While the existence of plane-wave states (\ref{plane_wave}) is a generic property
of any statistical model with broken U(1) symmetry, the period depends on the reference frame.
The Galilean transformation of the field $\Phi$ when going to the reference frame moving with the velocity ${\bf v}_0$
with respect to the original one,
\begin{equation}
\Phi({\bf r},t) \, \to\,  \Phi({\bf r},t)\,  - \, {{\bf v}_0 \cdot {\bf r} \over \gamma} ,
\label{Galilean}
\end{equation}
implies that the state wavevector changes to:
\begin{equation}
{\bf k} \, \to \, {\bf k} - {\bf v}_0 /\gamma .
\label{Galilean_k}
\end{equation}
Here $\gamma$ is the system-specific parameter relating the wavevector of the matter wave to the flow velocity.
In the quantum case, $\gamma=\hbar/m$, where $\hbar$ is the Planck's constant (in what follows we set it to unity)
and $m$ is the particle mass. This relativity of the period is quite unique for non-relativistic crystals!

What distinguishes superfluids from purely statistical models with broken U(1) symmetry is that broken U(1) symmetry
automatically entails breaking of the time-translation symmetry, and links superfluidity to yet another fundamental
phenomenon of {\it time crystallization} \cite{Wilczek,Shapere}.
Indeed, the phase $\Phi$ evolves in time in accordance with the universal Beliaev--Josephson--Anderson
relation (in the reference frame of the normal component)
\begin{equation}
\dot{\Phi} = -\mu .
\label{phi_dot}
\end{equation}
Here $\mu \equiv \mu_k$  is the chemical potential that depends on the wavevector of the superflow.
Equation (\ref{phi_dot}) readily follows from the generalised Gibbs distribution for a superfluid \cite{SBP}.
Its remarkable simplicity and universality is rooted in the fact that the phase $\Phi$ is canonically
conjugated to the total amount of matter
\begin{equation}
N = \int |\psi|^2 d^d r ,
\label{N}
\end{equation}
(the U(1) symmetry in question is the Noether's symmetry responsible for the conservation of $N$).
With the relation (\ref {phi_dot}), the expression (\ref{plane_wave}) can be upgraded to the formula
 \begin{equation}
\psi({\bf r},t)\, =\, \psi({\bf 0},0)\, e^{i {\bf k}\cdot {\bf r} -i\mu_k t} ,
\label{running_plane_wave}
\end{equation}
showing that, in the long-wave limit, the superfluid order parameter has the form of a running plane wave.
Hence, the superfluid state with a superflow is a space-time crystal, or, a time crystal in the of absence
of the superflow.

It is important to keep in mind that the value of  $\mu_k$ is relative. This is formally reminiscent of
(and even partially connected to) the relative nature of the wavevector ${\bf k}$: changing the reference
frame we change ${\bf k}$ and $\mu_k$. Furthermore, the chemical potential is defined only up to a global
constant prescribed by the convention about the ground state energy per particle.
In the non-relativistic physics, the rest energy of a free particle is typically set to zero.
The merely conventional character of this choice results in a certain constraint on the protocols
of measuring $\mu$ and the time-crystallization effect in superfluids, but does not exclude the effect
itself.

In their paper on the no-go theorem for equilibrium time crystals \cite{Watanabe}, Watanabe and Oshikawa
argued that it is the above-discussed relativity of the chemical potential that reconciles their theorem
with Eq.~(\ref{running_plane_wave}). However, the proof of the no-go theorem in Ref.~\cite{Watanabe} is
based on the implicit assumption that energy is the only additive constant of motion in a system  (cf. Ref.~\cite{Syrwid_2017}), which is
certainly not true for superfluids where (\ref{N}) is also a key constant of motion. The actual restriction
implied by the no-go theorem for equilibrium time crystals is two-fold: (i) An equilibrium time crystal is
supposed to have at least one additive constant of motion besides the energy. (ii) The observable
revealing the time crystallization has to violate the conservation of this constant.

In the light of the above discussion, it is instructive to identify a class of superfluid systems
that feature the effect of equilibrium space-/time-crystallization in the form of Eq.~(\ref{running_plane_wave})
while being free of the subtleties originating from the relative nature of ${\bf k}$ and ${\mu_k}$.
We observe that multicomponent (counterflow) superfluids belong to such a class. Here the quantity of interest,
$\Phi_{\rm ab}({\bf r},t)$, is the coarse-grained field of the phase difference between the components ``a" and ``b"
(the description stays exactly the same for arbitrary number of components, so we restrict ourselves to the
two-component case for simplicity). We further limit ourselves with counterflow superfluid states (see, e.g., \cite{SBP})
where the superfluid order exists exclusively in the field $\Phi_{\rm ab}({\bf r},t)$ but not in the individual phases
of the components. The long-wave equilibrium statistics of the two-component counterflow superfluid is isomorphic
to that of a single-component superfluid, rendering the system particularly simple and relevant for our purposes.
To exclude irrelevant long-wave degrees of freedom, we also assume that the normal component is pinned by either
disorder, walls, or an external periodic potential. The Beliaev--Josephson--Anderson
relation for $\Phi_{\rm ab}$ (its derivation from the Gibbs distribution is directly analogous to that in the
single-component case),
\begin{equation}
\dot{\Phi}_{\rm ab} = \mu_{\rm b}-\mu_{\rm a} ,
\label{phi_dot_ab}
\end{equation}
has the form of the Josephson relation for the standard ac Josephson effect between two single-component superfluids
(made of the same type of matter but having different chemical potentials). Similarly, the protocol of detecting the
rotation of the phase $\Phi_{\rm ab}$ can be based on simply creating a ``Josephson link" between components a and b.
In this regard, note that any protocol of revealing the time crystallization effect in the field $\Phi_{\rm ab}$ has to
deal with interactions explicitly violating the U(1)$\times$U(1) symmetry of the original system, which implies a process
converting components ``a" and ``b" into each other. The conceptual difference between this ``internal" Josephson effect
and its conventional counterpart is that now the frequency of the phase rotation---and thus the period of oscillations
of the ac Josephson current---is independent of the choice for counting energy in a single equilibrium sample.

In the presence of disorder or external periodic potential, the system has a natural reference frame. The absence of Galilean
invariance in this case does not yet mean that the period of the space crystal (\ref{plane_wave}) is not relative.
One can, in principle, design an experiment when this period is observed from a moving frame, in which case
Eqs.~(\ref{Galilean})--(\ref{Galilean_k}) still apply. However, in a counterflow superfluid with two components having
equal parameters $\gamma$, Galilean transformation (\ref{Galilean}) leaves the phase field intact,
\begin{equation}
\Phi_{\rm ab} \, \to\,  \Phi_{\rm ab}\,  -\, {{\bf v}_0 \cdot {\bf r} \over \gamma_a} \, +\, {{\bf v}_0 \cdot {\bf r} \over \gamma_b}
\, \equiv \, \Phi_{\rm ab} ,
\label{Galilean_ab}
\end{equation}
and the period of the corresponding space crystal is the same in any reference frame.

The difference between the single-component and counterflow superfluids becomes even more dramatic and instructive in the case of toroidal
geometry and rotating frame. The fictitious vector potential ${\bf A}_{\rm fict}$ emerging in the rotating frame brings about the gauge freedom. In the single-component superfluid in the rotating frame, the gauge freedom renders the notion of spatial phase difference ambiguous and thus ill defined. The gauge-invariant equivalent of the phase difference between the points ${\bf r}_1$ and  ${\bf r}_2$ has now the form of the line integral:
\begin{equation}
\int_{{\bf r}_1}^{{\bf r}_2} (\nabla \Phi - {\bf A}_{\rm fict})\cdot d{\bf l} .
\label{r1_r2}
\end{equation}
Even in the absence of topological defects in the field $\Phi$, this integral depends on the form of the line because of the term with ${\bf A}_{\rm fict}$. This, in particular, means that there is no experimental way of unambiguously measuring the phase difference in the rotating frame. In the case of counterflow superfluid, the counterpart of the integral (\ref{r1_r2}) reads [cf. Eq.~(\ref{Galilean_ab})]
\begin{equation}
\int_{{\bf r}_1}^{{\bf r}_2} (\nabla \Phi_{\rm ab} - {\bf A}_{\rm fict}+{\bf A}_{\rm fict})  \cdot d{\bf l} \, \equiv
\int_{{\bf r}_1}^{{\bf r}_2} \nabla \Phi_{\rm ab}  \cdot d{\bf l}.
\label{r1_r2_SCF}
\end{equation}
Now the phase difference is well defined and invariant with respect to the choice of the reference frame.

{\it  Experimental implementation 1: Cold atoms.} One possible realization of the counterflow superfluidity
is with multicomponent ultracold  bosons in optical lattices \cite{Kuklov_2003}.
A straightforward generalization of the protocol discussed in Ref.~\cite{PS_2018} (in the context of algebraic
time crystallization in a single-component two-dimensional superfluid) allows one to simultaneously study both
the space and time crystallization in such systems. The protocol is as follows.

$\bullet$ Consider a toroidal shape sample with close, but different, chemical potentials $\mu_a$ and $\mu_b$, in a state
with finite counterflow supercurrent. Introduce switchable internal Josephson links between the two components on two special
sites of the optical lattice separated by a distance ${\bf r}$.

$\bullet$  At time zero, turn the first link on for the duration $\Delta t$, such that ($\Delta \mu = |\mu_a  -  \mu_b|$)
\begin{equation}
 \Delta t \, \Delta \mu \, \ll \, 1 .
\label{Delta_t}
\end{equation}

$\bullet$  Keep both links switched off for a much longer time interval $t \, \Delta \mu \, > \, 1$ and then turn the second link on for the duration
$\Delta t$.

$\bullet$ Quickly, on time scales $\, \ll \, 1/\Delta \mu$, apply a deep optical lattice to localize all
          atoms in the system and count atom numbers $N_{\rm a}$ and $N_{\rm b}$ using single-site microscopy \cite{Bloch_2010,Greiner_2010}.

Repeating the protocol many times under identical conditions allows one to accumulate
representative statistics and process the data with the help of an auxiliary experimental run
that skips the next-to-last step of the above-described protocol.
The outcome of the auxiliary run is the expectation value $\bar{N}_{\rm ab} = \langle N_{\rm a} - N_{\rm b} \rangle$
that averages typical particle number differences taking place {\it right before} the two samples are
disconnected for a period of time $t$.  The key statistical observable is then \\
\begin{equation}
K(t) \, =\,  \langle \; [N_{\rm a}(t) - N_{\rm b}(t)  - \bar{N}_{\rm ab} ]^2  \, \rangle .
\label{K}
\end{equation}
In this expression, random particle number differences characterizing irreproducibility of the initial state preparation
cancel out and we are left with a signal reflecting spatial and temporal oscillations of the phase field
\begin{equation}
 \langle  \, \Phi_{\rm ab}({\bf r},t) - \Phi_{\rm ab}({\bf 0},0) \, \rangle .
\label{the_correlator}
\end{equation}
State preparation fluctuations are independent of ${\bf r}$ and $t$ and thus creates no problem except for that of a signal-to-noise ratio, which can be improved by collecting more statistics and optimizing setup parameters. To ensure that the space-/time-dependent contribution to
dispersion is large, one needs to have $J_0/\Delta \mu  \gg 1$, where $J_0$ is the Josephson constant (assumed to be the same for both links).

{\it  Experimental implementation 2: Bilayer superconductor.} A different---and interesting on its own---realization of
space-/time-crystallization effect in a counterflow superfluid is a bilayer superconducting annulus. When the thickness of the
two layers is small enough to suppress finite-temperature bulk superconductivity and tunneling between the layers is negligibly small,
the system still features (at appropriately low temperature) a two-dimensional neutral counterflow superfluid mode \cite{Babaev}.
In this case, the coarse-grained phase field $\Phi_{\rm ab} ({\bf r},t)$ describes the phase difference between the
layers ``a" and ``b." Because of the long-range current-current interaction between the layers via the vector potential,
the field $\Phi_{\rm ab} ({\bf r},t)$ remains algebraically ordered while the individual phase fields
$\Phi_{\rm a} ({\bf r},t)$ and $\Phi_{\rm b} ({\bf r},t)$ are destroyed at finite-temperature by the proliferation of vortices
that cost finite energy \cite{Babaev}.
The effect of surface superconductivity predicted recently by Samoilenka and Babaev \cite{Samoilenka} can be used to
create an interesting modification of the bilayer superconducting setup \cite{Babaev_private}, in which the two
layers are formed by adjacent surfaces of two superconducting materials that remain normal in the bulk.

The setup with switchable Josephson link(s) and subsequent counting of the electrons in each of the two layers appears to be impractical.
Instead, one can utilize the standard ac Josephson setup with one or two permanent links.
One link would be sufficient for revealing the time crystallization through the current-current correlation function \cite{PS_2018}.
To reveal both the space and time crystallization, one needs the second link (assume that both have the same Josephson constant $J_0$)
at a macroscopically large distance ${\bf r}$ from the first one. Operationally, the resulting device will behave as a hybrid of
an ac Josephson junction and a SQUID.
On the one hand, the it will be demonstrating the algebraic Josephson effect, see Ref.~\cite{PS_2018},
with the frequency prescribed by the superconducting analog of relation (\ref{phi_dot_ab}), where the chemical potential
difference $\mu_{\rm b}-\mu_{\rm a}$ is doubled because of the Cooper pairing.
On the other hand, the net amplitude of the Josephson current will depend on the phase shift ${\bf k} \cdot {\bf r}$ between the two junctions: \\
\begin{equation}
J_{\rm net} \propto J_0 \left| \cos {{\bf k}\! \cdot \! {\bf r}\over 2} \right| .
\label{SQUID}
\end{equation}
This way one can directly measure the projection of the wavevector ${\bf k}$ on the axis of ${\bf r}$.

{\it Conclusions and discussion.} Multicomponent superfluids---most notably, counterflow superfluids---unquestionably
feature the effects of space and time crystallization.  In this context, the space crystallization is understood broadly as the broken translation symmetry, irrespective of
its microscopic origin (including the role played by interactions) and relevant observables. The counterflow superfluidity was predicted theoretically some time ago but it has
not been yet realized in the lab. Observation of the space-/time-crystallization effects can be used for detecting this superfluid state
experimentally.

To comply with the no-go theorem \cite{Watanabe}, a system featuring the effect of time crystallization has to have an
additive conserved quantity different from energy, and measurements have to violate the conservation of this quantity.
The counterflow superfluids with potentially inter-convertible components satisfy these criteria; internal Josephson links
between the two components probe the order in the phase difference field through temporal correlations and spacial interference
of Josephson currents. The necessity for the time-crystallization probe to deal with the inter-conversion of the two components
explicitly follows from the Beliaev--Josephson--Anderson relation (\ref{phi_dot_ab}). Otherwise,
each of the two chemical potentials is defined up to its individual arbitrary additive constant reflecting the convention
about the counting zero for energy.

We discussed two different experimental setups and, correspondingly, two different protocols for the observation of
space-/time-crystallization in counterflow superfluids. The first setup, dealing with ultracold atoms in optical lattices,
appears to be the most universal and conceptually transparent. Here the probes are local in space and time and unquestionably
remove all concerns about non-equilibrium effects. Yet, an important aspect of realistic cold-atomic systems is that their sizes
are rather moderate. This makes them especially suitable for studying the finite-size effects leading to phase decoherence
and hence the finite linewidth of the Josephson effect power spectrum \cite{L_O,Kurkjian}. This setup is equally good for
detecting the genuine space and time crystallization in a three-dimensional system, as well as algebraic
space and time crystallization in lower dimensions.

The second experimental setup is based on a bilayer supeconductor with two spatially separated Josephson links between the layers to
study the effect of space crystallization via the interference of the two Josephson currents. The advantage of this setup is that
it allows one to employ standard experimental techniques for detecting Josephson effect in electronic systems. In particular, one can
use the emitted electromagnetic radiation to measure the frequency and the amplitude of the oscillating current.
This setup also appears to be natural for utilizing the space-/time-crystallization effects to reveal and study some other superfluid
phenomena and properties such as algebraic (as opposed to genuine) time crystallization \cite{PS_2018}, equilibrium statistics of
supercurrent states \cite{PS_2000}, and surface superconductivity \cite{Samoilenka}.  A minor shortcoming of this setup is that the system is two-dimensional so that it deals with the algebraic space and time crystallization.

It is instructive to put our results in a broader context of past and present activities addressing spontaneous
breaking of time-translation symmetry in equilibrium, steady state, and periodically driven (Floquet) systems.
At the moment, an exciting progress is being made---on both  theoretical and experimental side---with Floquet time crystals
(see, e.g., review \cite{Sacha_review} and references therein). By their very nature---the presence of a periodic drive---Floquet
time crystals break discrete time-translation symmetry as opposed to breaking continuous time-translation symmetry.
The discussed scenario for a macroscopic system to break continuous time-translation symmetry is most closely related to
the Kuramoto synchronization mechanism (see review \cite{Acebron} and references therein), when, under appropriate conditions,
local rotors get globally synchronized despite local fluctuations and disorder.

Our discussion was focused on the counterflow superfluid. Nevertheless, all the conclusions  apply to any
multicomponent superfluid because it inevitably has at least one counterflow mode.
In particular, a simple two-component Bose-Einstein condensate would be a reasonable system for applying
the above-mentioned protocol. In this regard, the manifestation of time crystallization in a two-component
Bose-Einstein condensate has been already observed in Ref.~\cite{Hall}. 
As opposed to our protocol of detecting the time-translation
symmetry breaking at equilibrium, the experiment of Ref.~\cite{Hall} starts by creating a coherent  non-equilibrium initial state with well-defined relative phase between the two components (by producing the second component out of the condensed first one). 
The evidence for the broken time-translation symmetry then comes in the form of long-lived oscillations of the 
relative phase of the two condensates.

If the two-pulse protocol for equilibrium states is modified to render the weak interconversion interaction global (uniform) 
rather than local, then the experiment would demonstrate---by the very fact that
a macroscopic equilibrium system features a finite response to such type of perturbation---a fundamental 
property of macroscopic time crystals: an inevitable presence
of long-range spatial correlations along with the oscillations in the time domain. 
This property can be interpreted in terms of time-dependent order parameter, see, e.g., Ref. \cite{Efetov}.

{\it Acknowledgements.}  We thank Egor Babaev, Frank Wilczek, Krzysztof Sacha,  Jon Machta, and David Hall for their interest and comments. This work was supported by the National Science Foundation under the grant DMR-1720465 and the MURI Program ``New Quantum Phases of Matter" from AFOSR.

\end{document}